\documentclass[aps,prl,twocolumn,groupedaddress,showpacs]{revtex4}
\usepackage{graphicx}
\usepackage{wasysym}
\draft

\begin{document}
\
\title{ Quantum Hall Effect in Bilayer Graphene: Disorder Effect
and Quantum Phase Transition}
\author{R. Ma$^{1,2}$, L. Sheng$^{3}$, R. Shen$^{1,3}$,
M. Liu$^{2}$ and D. N. Sheng$^1$}
\address{
$^1$Department of Physics and Astronomy, California State
University, Northridge, California 91330, USA\\
$^2$Department of Physics, Southeast University, Nanjing 210096, China \\
$^3$National Laboratory of Solid State Microstructures and
Department of Physics, Nanjing University, Nanjing 210093, China }

\begin{abstract}
We numerically study the quantum Hall effect (QHE) in bilayer
graphene based on tight-binding model in the presence of disorder.
Two distinct QHE regimes are identified in the full energy band
separated by a critical region with non-quantized Hall Effect. The
Hall conductivity around the band center (Dirac point) shows an
anomalous quantization proportional to the valley degeneracy, but
the $\nu=0$ plateau is markedly absent, which is in agreement with
experimental observation. In the presence of disorder, the Hall
plateaus can be destroyed through the float-up of extended levels
toward the band center and higher plateaus disappear first. The
central two plateaus around the band center are most robust against
disorder scattering, which is separated by a small critical region
in between near the Dirac point. The longitudinal conductance around
the Dirac point is shown to be nearly a constant in a range of
disorder strength, till the last two QHE plateaus completely
collapse.
\end{abstract}

 \maketitle

\section{I. Introduction}

Since the experimental discovery of an unusual half-integer quantum
Hall effect (QHE)~\cite{K. S. Novoselov1,Y. Zhang} in monolayer
graphene, the electronic transport properties of graphene related
materials have been extensively studied~\cite{J.Nilsson1,E.
McCann,J. Nilsson2,K. S. Novoselov2, J. G. Checkelsky, Y. Hasegawa,
D. A. Abanin, E. V. Gorbar, H. Min, E. V. Castro, Sheng}. Recently,
bilayer graphene
is found to show an anomalous behavior in its spectral and transport
properties,
which has attracted much experimental and theoretical
interest.
Theoretical studies~\cite{E. McCann,J. Nilsson2} show that interlayer coupling
modifies the intralayer relativistic spectrum to yield a
quasiparticle spectrum with a parabolic energy dispersion, which
implies that the quasiparticles in bilayer graphene cannot be
treated as massless but have a finite mass. Experiments have shown
that bilayer graphene exhibits an unconventional integer
QHE~\cite{K. S. Novoselov2}. The Landau level (LL) quantization
results in plateaus of Hall conductivity at integer positions
proportional to the valley degeneracy, but the plateau at zero
energy is markedly absent. The unconventional QHE behavior derives
from the coupling between the two graphene layers. The
quasiparticles in bilayer graphene are chiral and carry a Berry
phase 2$\pi$, which strongly affects their quantum dynamics.
However, a detailed theoretical understanding of the unconventional
properties of the QHE in bilayer graphene taking into account of the
full band structure and disorder effect is still lacking. As
established for a single layer graphene~\cite{Sheng} and
conventional quantum Hall systems~\cite{Sheng0}, the QHE phase
diagram in such a system is crucially depending on the topological
properties of the full energy band,  and thus can be naturally
determined in the  band model calculations.

In this work, we carry out a numerical study of the QHE in bilayer
graphene in the presence of disorder based upon a tight-binding
model. We reveal that the experimentally observed unconventional QHE
plateaus emerge near the band center, while the conventional QHE
plateaus appear near the band edges. The unconventional ones are
found to be much more stable to disorder scattering than the
conventional ones near the band edges. We further investigate the
quantum phase transition and obtain the phase boundaries $W_{c}$ for
different QHE states to insulator transition by calculating the
Thouless number~\cite{J.T.Edwards}. Our results show that the
unconventional QHE plateaus can be destroyed at strong disorder (or
weak magnetic field) through the float-up of extended levels toward
the band center and higher plateaus always disappear first. While
the $\nu =\pm 2$ QHE states are most stable,  the Dirac point at the
band center separating these two QHE states remains critical with a
nearly constant longitudinal conductance.

This paper is organized as follows.
In Sec.\ II, we introduce the model Hamiltonian. In Sec.\ III,
numerical results based on exact diagonalization and transport
calculations are presented. The final section contains a summary.

\section{II. The tight-binding model of bilayer graphene}

\begin{figure}[tbp]
\includegraphics[width=2.0in]{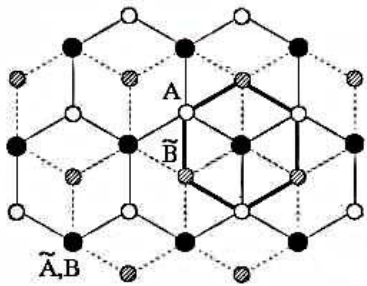}
\caption{(Color online) Schematic of bilayer graphene lattice with
AB (Bernal) stacking. Bonds in the bottom layer (A, B) are indicated
by solid lines and in the top layer ($\widetilde{A}$,
$\widetilde{B}$) by dash lines. A unit cell contains four atoms: A
(white circles), $\widetilde{B}$ (gray), $\widetilde{A}$B dimer
(solid).}
\end{figure}

We consider the bilayer graphene composed of two coupled hexagonal
lattice including inequivalent sublattices $A$, $B$ on the bottom
layer and $\widetilde{A}$, $\widetilde{B}$ on the top layer. The two
layers are arranged in the AB (Bernal) stacking~\cite{S. B.
Trickey,K. Yoshizawa}, as shown in Fig.\ 1, where  $B$ atoms are
located directly below $\widetilde{A}$ atoms, and $A$ atoms are the
centers of the hexagons in the other layer. The unit cell contains
four atoms $A$, $B$, $\widetilde{A}$, and $\widetilde{B}$, and the
Brillouin zone is identical with that of monolayer graphene. Here,
the in-plane nearest-neighbor hopping integral between $A$ and $B$
atoms or between $\widetilde{A}$ and $\widetilde{B}$ atoms is
denoted by $\gamma_{AB}
=\gamma_{\widetilde{A}\widetilde{B}}=\gamma_{0}$. For the interlayer
coupling, we take into account the largest hopping integral between
$B$ and $\widetilde{A}$ atoms $\gamma_{\widetilde{A}B}=\gamma_{1}$,
and the smaller hopping integral between $A$ and $\widetilde{B}$
atoms $\gamma_{A\widetilde{B}}=\gamma_{3}$. The values of these
hopping integrals are estimated to be $\gamma_{0}=3.16$ eV~\cite{W. W.
Toy}, $\gamma_{1}=0.39$ eV~\cite{A. Misu}, and $\gamma_{3}=0.315$
eV~\cite{R. E. Doezema}.

We assume that each monolayer graphene has totally $L_{y}$ zigzag
chains with $L_{x}$ atomic sites on each chain~\cite{Sheng}. The
size of the sample will be denoted as $N=L_{x}\times L_{y}\times
L_{z}$, where $L_{z}=2$ is the number of monolayer graphene planes
along the $z$ direction. In the presence of an applied magnetic
field perpendicular to the plane of the bilayer graphene, the
lattice model in real space can be written in the tight-binding
form:
\begin{eqnarray}
H&=&-\gamma_{0}\sum\limits_{\langle
ij\rangle}e^{ia_{ij}}(c_{i}^{\dagger }c_{j}+
\widetilde{c}_{i}^{\dagger }\widetilde{c}_{j}
)+ (-\gamma_{1}\sum\limits_{\langle
ij\rangle_1}e^{ia_{ij}}c_{jB}^{\dagger
}\widetilde{c}_{i\widetilde{A}}\nonumber\\
&-&\gamma_{3}\sum\limits_{\langle
ij\rangle_3}e^{ia_{ij}}c_{iA}^{\dagger
}\widetilde{c}_{j\widetilde{B}}
+h.c.)+\sum\limits_{i}w_{i}(c_{i}^{\dagger
}c_{i}+\widetilde{c}_{i}^{\dagger }\widetilde{c}_{i}),
\end{eqnarray}
where $c_{i}^{\dagger}$($c_{iA}^{\dagger}$),
$c_{j}^{\dagger}$($c_{jB}^{\dagger}$) are creating operators on $A$
and $B$ sublattices in the bottom layer, and
$\widetilde{c}_{i}^{\dagger}$($\widetilde{c}_{i\widetilde{A}}^{\dagger}$),
$\widetilde{c}_{j}^{\dagger}$($\widetilde{c}_{j\widetilde{B}}^{\dagger}$)
are creating operators on $\widetilde{A}$ and $\widetilde{B}$
sublattices in the top layer. The sum $\sum_{\langle ij\rangle}$
denotes the intralayer nearest-neighbor hopping in both layers,
$\sum_{\langle ij\rangle_1}$ stands for interlayer hopping between
the $B$ sublattice in the bottom layer and the $\widetilde{A}$
sublattice in the top layer, and $\sum_{\langle ij\rangle_3}$ stands
for the interlayer hopping between the $A$ sublattice in the bottom
layer and the $\widetilde{B}$ sublattice in the top layer, as
described above. $w_{i}$ is a random disorder potential uniformly
distributed in the interval $w_{i}\in \lbrack -W/2,W/2]\gamma_0$.
The magnetic flux per hexagon $\phi =\sum_{{\small
{\mbox{\hexagon}}}}a_{ij}=%
\frac{2\pi }{M}$, with $M$ an integer. The total flux through the
sample is $N\frac {\phi}{2\pi}$, where $N=L_{x}L_{y}/M$ is taken to
be an integer.
When $M$ is commensurate with $L_x$ or $L_y$,
the magnetic periodic boundary conditions are reduced to the
ordinary periodic boundary conditions.

\section{III. Results and Discussion}

The eigenstates $\vert\alpha\rangle$ and eigenenergies
$\epsilon_\alpha$ of the system are obtained through exact
diagonalization of the Hamiltonian Eq.\ (1), and the Hall
conductivity $\sigma _{xy}$ is calculated by using the Kubo formula
\[
\sigma _{xy}= \frac{ie^{2}\hbar}{S}\sum_{\alpha, \beta}\frac{\langle
\alpha\mid V_x\mid\beta\rangle\langle\beta\mid V_y\mid\alpha
\rangle-h.c.}{(\epsilon_\alpha-\epsilon_\beta)^2},
\]
where $S$ is the area of the sample, $V_{x}$ and $V_{y}$ are the
velocity operators. In Fig.\ 2a, the Hall conductivity $\sigma _{xy}
$ and electron density of states are plotted as functions of
electron Fermi energy $E_{f}$ for a clean sample ($W=0$) at system
size $N=96\times 24\times 2$ with magnetic flux $\phi =\frac{2\pi
}{48}$, which illustrates the overall picture of the QHE in the full
energy band. From the electron density of states, we can see the
discrete LLs. We will call  central LL at $E_f=0$ the $n=0$ LL, the
one just above (below) it the $n=1$ ($n=-1$) LL, and so on.
According to the behavior of $\sigma _{xy}$, the energy band is
naturally divided into three different regimes. Around the band
center, the Hall conductivity is quantized as $\sigma _{xy}=\nu
\frac{e^{2}}{h}$, where $\nu =kg_{s}$ with $k$ an integer and
$g_{s}=2$ for each LL due to double-valley degeneracy~\cite{E.
McCann,Sheng} (the spin degeneracy will contribute an additional
factor $2$, which is omitted here). With each additional LL being
occupied, the total Hall conductivity is increased by
$g_{s}\frac{e^{2}}{h}$. This is an invariant as long as the states
between the $n$-th and $(n-1)$-th LL are localized. $\sigma _{xy}=0$
at the particle-hole symmetric point $E_{f}=0$, which corresponds to
the half-filling of the central LL. However, there is no
$\sigma_{xy}=0$ quantized Hall plateau. These anomalously quantized
Hall plateaus agree with the results observed experimentally in
bilayer graphene~\cite{K. S. Novoselov2}.

The Hall conductivity near the band edges, however, is quantized as
$\sigma _{xy}=k \frac{e^{2}}{h}$ with $k$ an integer, as in the
conventional QHE systems. Remarkably, around $E_{f}=\pm \gamma_0 $
(within a narrow energy region $\Delta E\sim 0.4\gamma_0$), there
are two critical regions which separate the unconventional and
conventional QHE states, where the Hall conductance quantization
is lost. These crossover regions also correspond to a novel
transport regime, where the Hall resistance changes sign and the
longitudinal conductivity exhibits metallic behavior.
 The singular behavior of the Hall conductivity in the crossover regions
is likely to originate from the Van Hove singularity in the electron
density of states at $B=0$ limit.
In Fig.\ 2b, the quantization
rule of the Hall conductivity in this unconventional region for
three different strengths of magnetic flux is shown. With decreasing
magnetic flux from $\protect\phi =\frac{2\protect\pi }{12}$ to
$\frac{2\protect\pi }{48}$, more quantized Hall plateaus emerge
following the same quantization rule as the gap between the LLs is
reduced.


\begin{figure}[tbp]
\includegraphics[width=3.3in]{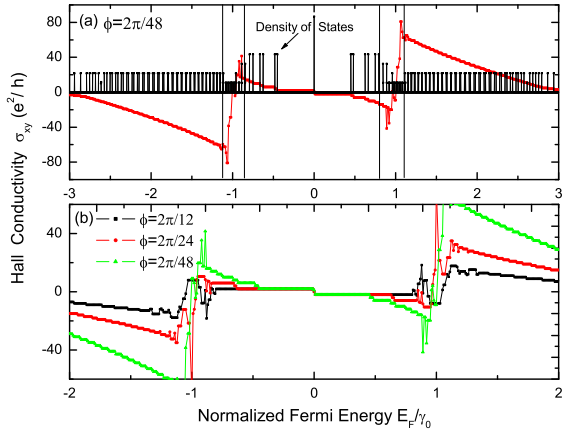}
\caption{(Color online) (a) Calculated Hall conductivity and
electron density of states in the full energy band for magnetic flux
$\protect\phi =\frac{2\protect\pi }{48}$ or M=48, and (b) the Hall
conductivity near the band center for $\protect\phi
=\frac{2\protect\pi }{12}$, $\frac{2\protect\pi }{24}$ and
$\frac{2\protect\pi }{48}$. The disorder strength is set to $W=0$
and $N=96\times 24\times 2$ in all cases. Inset: Hall conductivity
at the band center. Here, the spin degrees of freedom are omitted,
so $g_{s}=2$ and $g_{s}=1$ for the unconventional and conventional
regions, respectively.}
\end{figure}

\begin{figure}[tbp]
\includegraphics[width=3.0in]{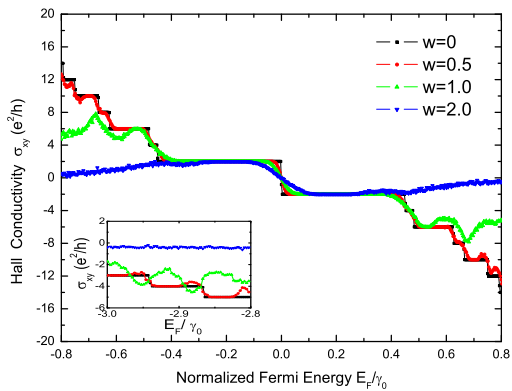}
\caption{(Color online) Unconventional Hall conductivity as a
function of electron Fermi energy near the band center for four
different disorder strengths each averaged over $400$ disorder
configurations. Inset: conventional Hall conductivity near the lower
band edge. Here, $\protect\phi =\frac{2\protect\pi }{48}$ and the
sample size is $N=96\times 24\times 2$.}
\end{figure}

Now we study the effect of random disorder on the unconventional QHE
in bilayer graphene. In Fig.\ 3, the Hall conductivity around the
band center is shown as a function of $E_{f}$ for four different
disorder strengths at system size $N=96\times 24\times 2$ with
magnetic flux $\phi =\frac{2\pi }{48}$.  We can see that the
plateaus with $\nu =\pm 10,\pm 6$ and $\pm 2$ remain well quantized
at $W=0.5$. We mention that the $\nu=\pm 4,\pm 8$ plateaus are
unclear at this relatively weak disorder strength because of very
small plateau widths and relatively large localization lengths (the
critical $W_c$  for each plateau will be  obtained based on our
larger size calculations of the Thouless number as presented later).
With increasing $W$, higher Hall plateaus (with larger $|\nu |$) are
destroyed first. At $W=2.0$, only the $\nu =\pm 2$ QHE remain
robust.
The last two plateaus $\nu =\pm 2$ eventually disappear around
$W\sim 3.2$. For comparison, the QHE near the lower band edge is
shown in the inset, where all plateaus disappear at a much weaker
disorder strength $W\geq 1.0$. This clearly indicates that under the
same conditions, the unconventional QHE around the band center is
much more stable than the conventional QHE near the band edges.
Clearly, after the destruction of the conventional QHE states near
the band edge, these states become localized. Then the topological
Chern numbers initially carried by these states will move towards
band center in a similar manner to the single-layer graphene
case~\cite{Sheng}. Thus we observe that the destruction of the
unconventional QHE states near the band center is due to the
float-up of extended levels.

\begin{figure}[tbp]
\includegraphics[width=3.3in]{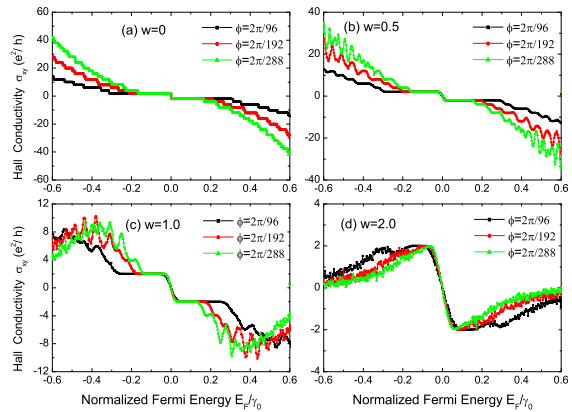}
\caption{(Color online) Calculated Hall conductivity with
weaker magnetic flux $\protect\phi =\frac{2\protect\pi }{96}$,
$\frac{2\protect\pi }{192}$ and $\frac{2\protect\pi }{288}$ for four
different disorder strengths each averaged over $400$ disorder
configurations. Here, the sample size is $N=96\times 24\times 2$.}
\end{figure}

To study the fate of the IQHE at weak magnetic field limit, we
reduce the strength of magnetic field.
In Fig.\ 4, the Hall conductivities around the band center with
weaker magnetic flux $\phi =\frac{2\pi }{96}$, $\frac{2\protect\pi
}{192}$ and $\frac{2\protect\pi }{288}$ are shown for different
disorder strengths and system size $N=96\times 24\times 2$.
In Fig.\ 4a, a lot more  well quantized Hall plateaus emerge for a clean
sample($W=0$), if we  compare them with the results in Fig.\ 2b.
In Fig.\ 4b, 4c and 4d, we can see that with the increasing of
the disorder strength $W$, Hall
plateaus are destroyed faster for the system with
  weaker magnetic flux $\phi$. At
$W=2.0$, the most robust Hall plateaus at $\nu =\pm 2$ remain well
quantized for magnetic flux $\phi =\frac{2\pi }{96}$ and
$\frac{2\protect\pi }{192}$, however, they already disappear for
weaker magnetic flux $\phi =\frac{2\pi }{288}$. Our flux $2\pi/M$ in
each hexagon the magnetic field is $B \sim 1.3\times 10^5/M$
Tesla\cite{weakb}. Thus the weakest $B$ we used is about $451$
Tesla. This is a very large magnetic field comparing to the
experimental ones around $B\sim 40 $ Tesla. However, the topology of
the QHE and how they disappear with the increase of the disorder
strength $W$ remain to be the same as the stronger $B$ cases as
demonstrated in Fig. 4a-4d. Thus, we establish that the obtained
behavior of QHE for bilayer graphene will survive at weak $B$ limit.

We further investigate the quantum phase transition of the bilayer
graphene electron system. In order to determine the critical
disorder strength $W_{c}$ for the different QHE states, the Thouless
number $g$ is calculated by using the following
formula~\cite{J.T.Edwards},
\[
g= \frac{\Delta E}{dE/dN}\ .            
\]
Here, $\Delta E$ is the geometric mean of the shift in the energy
levels of the system caused by replacing periodic by antiperiodic
boundary conditions, and $dE/dN$ is the mean spacing of the energy
levels. The Thouless number $g$ is proportional to the longitudinal
conductance $G$. In Fig.\ 5, we show some examples of calculated
Thouless number for a relatively weak flux $\phi =\frac{2\pi }{48}$
and some different disorder strengths to explain how quantum phase
transitions and the related phase boundaries $W_c$ are determined.
In Fig.\ 5a, the calculated Thouless number $g$ and Hall
conductivity $\sigma _{xy}$ as a function of $E_{f}$ at a weak
disorder strength $W=0.2$ are plotted. Clearly, each valley in
Thouless number corresponds to a Hall plateau and each peak
corresponds to a critical point  between two neighboring Hall
plateaus. We can also call the first valley just above (below)
$E_f=0$ the $\nu=-2$ ($\nu=2$) QHE state, the second one the
$\nu=-4$ ($\nu=4$) state, and so on, as same as the Hall plateaus.
In Fig.\ 5b-5d, we see that with increasing $W$, higher QHE states
(valleys) are destroyed first. At $W=W_c=1.0$ (see Fig.\ 5b), the
valleys with $\nu=\pm 12$ disappear, which correspond to the
destruction of the $\nu =\pm 12$ Hall plateau states. Therefore,
$W_c=1.0$ is the critical disorder strength, at which the $\nu =\pm
12$ plateau states change to an insulating phase. At $W=W_c=1.3$
(see Fig.\ 5c), the valleys with $\nu =\pm 8$ disappear, which
indicates the destruction of the $\nu =\pm 8$ QHE states and their
transition into the insulating phase. When $W=W_{c}=3.2$ (see Fig.\
5d), the most stable QHE states with $\nu=\pm 2$ eventually
disappear, which indicates all QHE phases are destroyed by disorder.
All the phase boundaries $W_{c}$ between the different QHE states
are determined in the same manner and tabulated in Table 1.

\begin{figure}[tbp]
\par
\includegraphics[width=3.3in]{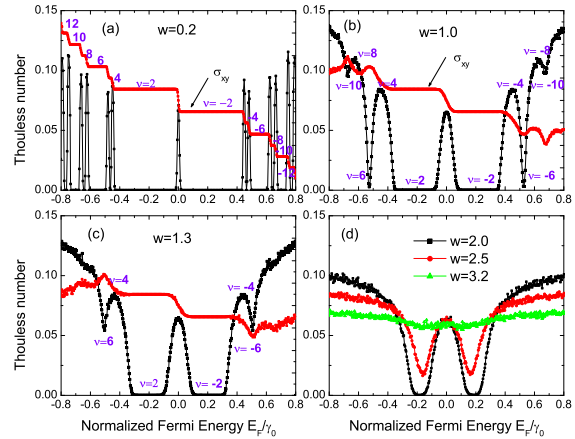}
\caption{(Color online) (a)-(c) Calculated Thouless number and Hall
conductivity for three different disorder strengths, and (d)
Thouless number for other three disorder strengths, each data point
being averaged over $400$ disorder configurations. Here,
$\protect\phi =\frac{2\protect\pi }{48}$ and the sample size are
taken to be $N=96\times 48\times 2$ and $N=96\times 24\times 2$ in
the calculations of Thouless number and Hall conductivity,
respectively.}
\end{figure}

We now focus on the region around $E_{f}=0$. In Fig.\ 6, we show the
Thouless number for some different disorder strengths at system size
$N=96\times 24\times 2$ and magnetic flux $\phi =\frac{2\pi }{48}$.
We can see that the Thouless number shows a central peak at
$E_{f}=0$. With increasing the disorder strength, the width of the
peak increases and its height remains nearly unchanged. This
behavior may suggest an interesting effect that the extended states
originally sited at the critical point $E_f=0$ splits in the
presence of disorder. However, the splitting is too small to induce
two separated peaks in the Thouless number for the present sample
sizes we can approach. Instead, it leads to a widened peak of
unreduced height.
This behavior also indicates that the critical longitudinal
conductance in a small finite region near $E_f=0$ is almost constant
about $2e^2/h$ according to the proportionality of Thouless number to
longitudinal conductance. We have also confirmed this conclusion by
direct Kubo formula calculation, in which the system size that can be
approached is however much smaller.

\begin{table}[htb]
\begin{tabular}{@{}cc}\hline
 Hall plateaus index   & critical point $W_c$ \\
  \hline
$\nu=\pm 12$&1.0\\
$\nu=\pm 10$&1.2$\pm 0.1$\\
$\nu=\pm 8$&1.3$\pm 0.1$\\
$\nu=\pm 6$&1.6$\pm 0.1$\\
$\nu=\pm 4$&1.7$\pm 0.1$\\
$\nu=\pm 2$&3.2\\
\hline
\end{tabular}
\caption{The phase boundaries $W_{c}$ for the different Hall
plateaus.}
\end{table}

\begin{figure}[tbp]
\par
\includegraphics[width=3.4in]{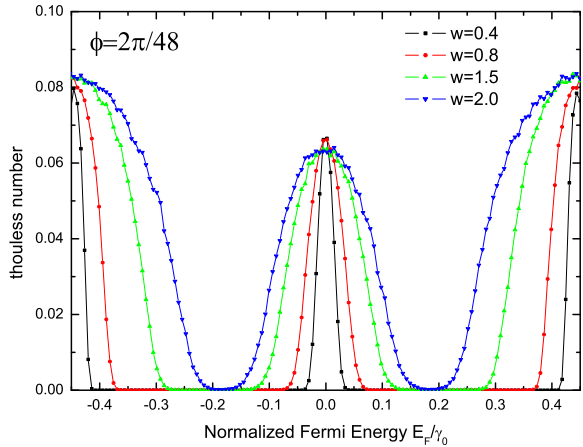}
\caption{(Color online) (a) Thouless number for five different
disorder strength, each point being averaged over $400$ disorder configurations.
Here, $\protect\phi =\frac{2\protect\pi }{48}$ and the sample size is
$N=96\times 48\times 2$.}
\end{figure}

\section{IV. Summary}

In summary, we have numerically investigated the QHE in bilayer
graphene based on tight-binding model in the presence of disorder.
The experimentally observed unconventional QHE is reproduced near
the band center. The unconventional QHE plateaus around the band
center are found to be much more stable than the conventional ones
near the band edges. Our results of quantum phase transition
indicate that with increasing disorder strength, the Hall plateaus
can be destroyed through the float-up of extended levels toward the
band center and higher plateaus always disappear first. At
$W=W_{c}=3.2$, the most stable QHE states with $\nu=\pm 2$
eventually disappear, which indicates transition of all QHE phases
into the insulating phase. A small critical region is observed
between the $\nu=\pm 2$ plateaus, where the longitudinal conductance
remains almost constant about $2e^2/h$ in the presence of moderate
disorder, possibly due to the splitting of the critical point
originally sited at $E_f=0$. We mention that in our numerical
calculations, the magnetic field is much stronger than the ones one
can realize in the experimental situation, as limited by current
computational ability. However, the phase diagram we obtained is
robust and applicable to weak field limit since it is determined by
the topological property of the energy band as clearly established
for single layer graphene~\cite{Sheng} and conventional quantum Hall
systems~\cite{Sheng0}. We further point out that the continuum model
can not be used to address the fate of the quantum Hall effect in
strong disorder or weak magnetic field limit. Because in such a
model, both the band bottom and band edge are pushed to infinite
energy limit, and thus one will not be able to see the important
physics of opposite Chern numbers annihilating each other to destroy
the IQHE\cite{Sheng}.

\textbf{Acknowledgment:} This work is supported by the US DOE grant
DE-FG02-06ER46305 (RS, DNS) and the NSF grant DMR-0605696 (RM, DNS).
We thank the KITP for partial support through the NSF grant
PHY05-51164. We also thank the partial support from the State
Scholarship Fund from the China Scholarship Council, the Scientific
Research Foundation of Graduate School of Southeast University of
China (RM), the National Basic Research Program of China under grant
Nos.: 2007CB925104 and 2009CB929504 (LS), and the NSF of China grant
Nos.: 10874066 (LS), 10504011 (RS), 10574021 (ML), the doctoral
foundation of Chinese Universities under grant No. 20060286044(ML).

\end{document}